\documentclass[preprint,12pt]{elsarticle}

\usepackage{amssymb}

\usepackage{amsmath,amsfonts}
\usepackage{algorithmic}
\usepackage{algorithm}
\usepackage{array}
\usepackage{textcomp}
\usepackage{url}
\usepackage{verbatim}
\usepackage{graphicx}
\usepackage{cite}
\usepackage{mathtools}
\usepackage{cases}
\usepackage[table,xcdraw]{xcolor}
\usepackage{upgreek}
\usepackage{caption}
\usepackage{listings}
\usepackage{fancyvrb}
\usepackage{ifthen}
\usepackage{soul}

\definecolor{myazul}{rgb}{0,0.4431,0.7373}
\definecolor{myazul2}{rgb}{0,0.25,0.75}
\definecolor{mynaranja}{rgb}{0.8471,0.3216,0.0941}
\definecolor{myverde}{rgb}{0.47,0.67,0.19}
\definecolor{myverde2}{rgb}{0.0,0.5,0.0}
\definecolor{myvioleta}{rgb}{0.4,0.18,0.56}
\definecolor{myvioleta2}{rgb}{0.4,0.18,0.86}
\definecolor{myceleste}{rgb}{0.6,0.85,0.93}
\definecolor{myamarillo}{rgb}{0.93,0.88,0.17}
\definecolor{myrojo}{rgb}{0.84,0.04,0.00}
\definecolor{myrojo2}{rgb}{0.71,0 ,0}
\definecolor{mynegro}{rgb}{0.01,0.01,0.01}
\definecolor{backgroundColour}{rgb}{0.95,0.95,0.92}

\newcommand{\abs}[1]{\left|#1\right|}
\newcommand{\comillas}[1]{``#1''}
\newcommand{\micron}{~\upmu\mathrm{m}}

\newcommand{\deff}{d_{\mathrm{eff}}}
\newcommand{\lcr}{L_{\mathrm{cr}}}

\newcommand{\trt}{t_\mathrm{rt}}

\newcommand{\dk}{\Delta k}
\newcommand{\fpm}{f_{\mathrm{pm}}}

\newcommand{\fourier}[1]{\mathcal{F}\left\lbrace #1 \right\rbrace}
\newcommand{\invfourier}[1]{\mathcal{F}^{-1}\left\lbrace #1 \right\rbrace}

\newcommand{\Pin}{P_{\mathrm{in}}}
\newcommand{\Pth}{P_{\mathrm{th}}}

\newboolean{EDITEDCOLOR}
\setboolean{EDITEDCOLOR}{false} 
\newcommand{\edited}[1]{%
    \ifthenelse{\boolean{EDITEDCOLOR}}%
        {\textbf{\textcolor{myrojo2}{#1}}}%
        {\textcolor{black}{#1}}%
}

\newcounter{bla}

\journal{Computer Physics Communications}

\begin{document}

\begin{frontmatter}



\title{CUDA-based optical parametric oscillator simulator}


\author[a]{A. D. Sanchez\corref{cor1}}
\cortext[cor1] {Corresponding author.\\\textit{E-mail address:} alfredo.sanchez@icfo.eu}
\author[b]{S. Chaitanya Kumar}
\author[a,c]{M. Ebrahim-Zadeh}

\address[a]{ICFO-Institut de Ciencies Fotoniques, Mediterranean Technology Park, 08860 Castelldefels, Barcelona, Spain.}
\address[b]{Tata Institute of Fundamental Research Hyderabad, 36/P Gopanpally, Hyderabad 500046, Telangana, India.}
\address[c]{Instituciò Catalana de Recerca i Estudis Avancats (ICREA), Passeig Lluis Companys 23, Barcelona 08010, Spain.}

\begin{abstract}
The coupled-wave equations (CWEs) in nonlinear optics are the fundamental starting point in the study, analysis, and understanding of various frequency conversion processes in dielectric media subjected to intense laser radiation. In this work, a useful package for the modeling of optical parametric oscillators (OPOs) based on the Split-Step Fourier Method algorithm is presented. The algorithm is scripted in the CUDA programming language in order to speed up the calculations and obtain results in a relatively short time frame by using a graphics processing unit (GPU). Our results show a speedup higher than 50X for vector size of $2^{14}$ in comparison with the analogous code scripted for running only in CPU. The package implements the CWEs to model the propagation of light in second-order nonlinear crystals widely used in optical frequency conversion experiments. In addition, the code allows the user to adapt the cavity configuration by selecting the resonant electric fields and/or incorporating intracavity elements. The package is useful for modeling OPOs or other mathematically similar problems.
\end{abstract}


\end{frontmatter}



{\bf PROGRAM SUMMARY/NEW VERSION PROGRAM SUMMARY}

\begin{small}
\noindent
{\em Program Title: \verb|cuOPO|}                                          \\
{\em CPC Library link to program files:} (to be added by Technical Editor) \\
{\em Developer's repository link: https://github.com/alfredos84/cuOPO} \\
{\em Code Ocean capsule:} (to be added by Technical Editor)\\
{\em Licensing provisions: MIT}  \\
{\em Programming language: CUDA}                                   \\
{\em Supplementary material:}                                 \\
{\em Journal reference of previous version:}*                  \\
{\em Does the new version supersede the previous version?:}*   \\
{\em Reasons for the new version:*}\\
{\em Summary of revisions:}*\\
{\em Nature of problem:}\\
   The problem that is solved in this work is that of two or three coupled differential equations that describe the propagation of light in a second order nonlinear medium, allowing the three-wave mixing process. By placing the medium in an optical cavity, an optical parametric oscillator is formed. The optical cavity is modeled by including the appropriate boundary conditions for the differential equations. As a result we obtain the electric fields of the interacting waves in the time and frequency domains.\\
{\em Solution method: }\\
  The coupled differential equations are solved using the well-known fixed-step split-step Fourier method. Due to the eventual computational demand that some problems may have, we chose to implement the coupled equations in the CUDA programming language. This allows us to significantly speed up simulations, thanks to the computing power provided by a graphics processing unit (GPU) card. The output files obtained are the interacting electric fields, which have to be analyzed during post-processing.\\
{\em Additional comments including restrictions and unusual features (approx. 50-250 words):}\\
   \\

\end{small}

\section{Introduction}
\label{sec:intro}
\noindent
Nonlinear optical processes have been a cornerstone of laser science and technology for over sixty years. Nonlinear optics has had tremendous impact in photonics, enabling unprecedented advances in multitude of disciplines in fundamental and applied physics, communications, quantum technologies, biology, medicine, and more. The coupled-wave equations (CWEs) are the backbone of nonlinear optics~\citep{armstrong1962interactions} and fundamental to the description of frequency conversion phenomena involving the interaction of optical waves in a dielectric medium. Fundamental to the understanding of nonlinear processes are the CWEs, which can be regarded as equivalent to Maxwell's wave equations in linear optics. CWEs are a set of nonlinear partial differential equations, related by a so-called \comillas{coupling term}, which well describes wave mixing in such media. Of the various nonlinear interactions, three-wave mixing (TWM) processes in non-centrosymmetric media exhibiting second-order nonlinear polarization are of considerable interest because of their ability to generate coherent radiation in new spectral and temporal regions not accessible to conventional lasers. Such TWM processes include sum- or difference-frequency-generation (SFG or DFG), second-harmonic-generation (SHG), optical parametric generation (OPG), and the optical parametric oscillator (OPO)~\citep{shen1984principles, boyd2020nonlinear}. 

In most cases, CWEs are solved numerically since analytical solutions are generally not available. Consequently, numerical algorithms are used to provide solutions quickly and efficiently. The well-known split-step Fourier method~\citep{agrawal2000nonlinear} (SSFM) was widely used to solve the CWEs~\citep{smith2003degenerate, sabouri2013thermal} in second-order media because of its higher speed compared to others algorithms~\citep{taha1984analytical}. 

It is common in literature to find works where the CWEs are solved using homemade codes adapted to their own requirements~\citep{mosca2018modulation,roy2021spectral}. On the other hand, there is a user-friendly software available which provides several tools to design, analyze and optimize OPOs using a wide database of nonlinear crystal~\citep{smith2015introduction}. This software has been successfully used in different scenarios, for example, to model OPOs in nanosecond pulsed  configurations~\citep{smith1995comparison,smith1999numerical}. 

In this work, we provide a useful new computational tool for modeling OPOs in practical and experimentally realizable format based on the solution of CWEs. Since CWEs belong to the family of parallelizable problems, we take advantage of the hardware provided by the GPUs to speed up our calculations, especially when massive and sequential simulations are required in a short time frame using a commercial desktop computer. Our results show that the achieved speedup is approximately 50X compared to analogous code adapted to run only on CPU. In practical terms, a standard simulation, e.g. $2\times10^4$ round-trips, using our code can take just under 20 minutes, while on CPU around 30 hours.
We present an implementation of CWEs based on the so-called \textit{symmetrized} split-step Fourier Method, whose error of $\mathcal{O}(dz^3)$, where $dz$ is the step size along the direction of propagation, has been previously analyzed~\citep{arisholm1997general}. Our implementation uses a fixed step size along the electric field  direction within the nonlinear medium. However, if required, the user can modify the code according to the specific needs of each problem, for instance following the analysis performed in Refs.~\citep{BALAC20141,comparisonSSFM2014}. With the constantly growing demand and increasing interest in the study and design of nonlinear optical sources, and in particular novel OPO devices for practical applications as well as fundamental studies~\citep{Sanchez2022Ultrashort}, this tool is very useful to analyze novel system architectures that are unique. Such simulations in the continuous-wave cw regime were previously limited by the long computational time, which is here alleviated by using the GPU in combination with our package. Although the present package simulates second-order processes, adding terms to the model, namely the Kerr effect or higher-order dispersion terms, should not be difficult for any user and its implementation would require minimal effort. One example is the coupled nonlinear Schr\"odinger equations (CNSE), which model the interaction in cubic media, such as optical fibers~\citep{caplan2013nlsemagic,ferreira2019high}. This type of scheme can be easily adapted to our code, either to combine quadratic with cubic media in a resonant cavity or to adapt our code to a specific single four-wave mixing problem. We would like to emphasise that the real advantage of the GPU together with the SSFM for solving CWEs could be exploited while simulating resonant cavities involving multiple round-trips, where the dynamic evolution of the intensities and phase of the interacting waves can be studies. Such a task is time-consuming using a CPU, particularly while simulating resonators involving cw fields with low gain, requiring significantly higher number of round-trips before the cw OPO can breach threshold, as compared to a pulsed OPO. The flexible parameter space, ability to incorporate active optical components and capture their time evolution leads to exciting insights in to various fundamental aspects of a new class of parametric systems such as self-phase-locking~\citep{roy2021spectral, Nandy20}, novel devices for ultrashort pulse generation~\citep{Sanchez2022Ultrashort} and optical solitons in OPOs~\citep{ODonnell20}.

The algorithm used here has been developed in the framework of bulk OPOs, in other words oscillators with discrete components. However, the model is extendable to integrated photonic systems~\citep{Sanchez2022Ultrashort}, which could imply minimal changes in the code, if the CWEs are to be used to model the system.

\section{Coupled-wave equations with boundary conditions and intracavity elements}
\label{sec:model}
\noindent

\subsection{Theoretical model}
\label{sec:theory}
\begin{figure}[htbp]
    \centering
    \includegraphics[width=1.0\textwidth]{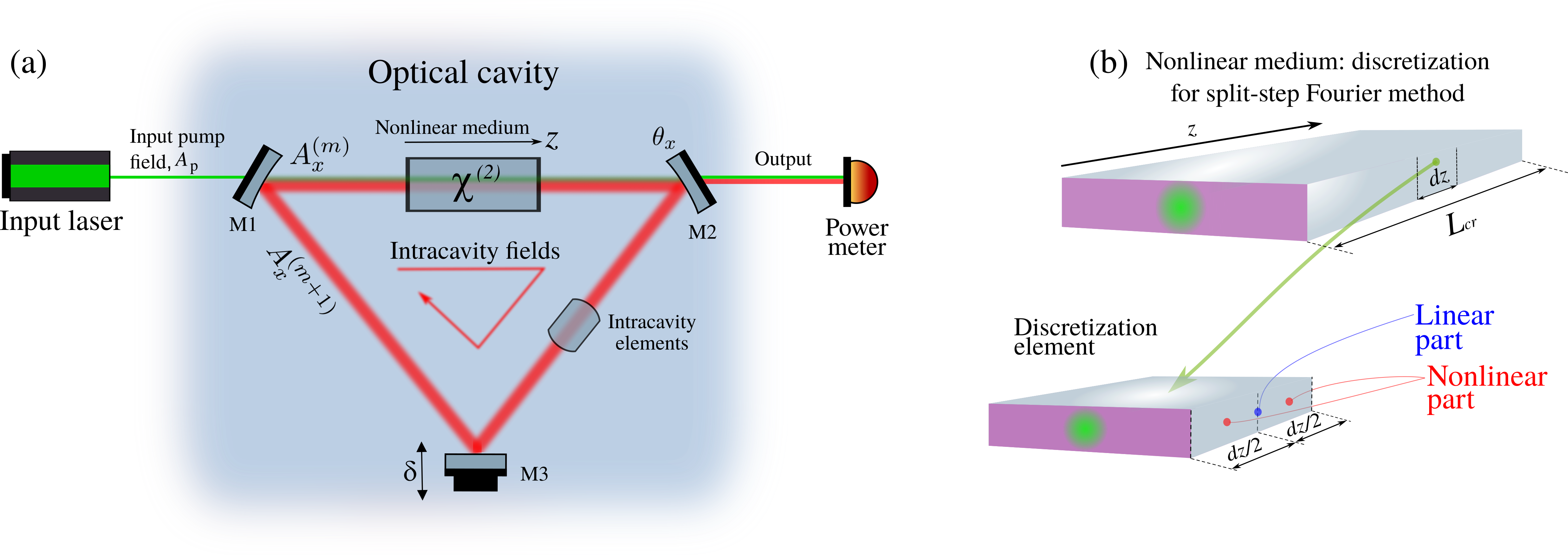}
    \caption{(a) Schematic of the OPO configuration used in our modelling using this package. (b) Crystal discretization depiction in the symmetrized SSFM.}
    \label{fig:scheme}
\end{figure}

\noindent
The scientific problem solved by this software is that of the CWEs governing the TWM processes of OPG in a second-order dielectric crystal placed within an optical resonator. With the crystal enclosed in an optical cavity, the process is referred to as optical parametric oscillation and the device is known as an OPO. The oscillator is formed in a ring cavity incorporating mirrors that provide resonance at the optical waves generated through the OPG process~\citep{ebrahimzadeh2001optical}. Figure~\ref{fig:scheme}(a) schematically depicts the basic configuration of the OPO that our code can simulate. It comprises an input laser that serves as a pump field, $A_p$, incident on a ring cavity formed by mirrors, M1-M3, a second-order nonlinear gain medium, $\chi^{(2)}$, which is typically a transparent non-centrosymmetric dielectric crystal that allows the generation of two optical waves, signal and idler, at new wavelengths through the OPG process. In the examples shown in this work we deploy a periodically-poled lithium niobate crystal (PPLN) as the nonlinear gain medium to show how the package performs. There are other intracavity elements that users can incorporate, such as an electro-optic modulator (EOM) or etalon which can be mathematically modelled~\citep{Sanchez2022Ultrashort}. As can be seen from Figure~\ref{fig:scheme}(a), M2 acts as an output coupler with power transmittance, $\theta_x$, while the mirror M3 is allowed to move for precise cavity length adjustment enabling cavity detuning, $\delta$. The superscript, $m$, stands for the $m^{\mathrm{th}}$ round-trip, while the subscript, $x$, stands for pump, signal or idler ($p,s$~or~$i$) intracavity electric fields, respectively. 

Each of the CWEs compute the evolution of an electric field at a specific wavelength, which interacts with the other two fields. In the three-wave parametric process, the interacting fields are pump, signal, and idler, each at angular frequencies, $\omega_p$, $\omega_s$, and $\omega_i$, respectively. In such a TWM process a photon at $\omega_p$ is annihilated, and two photons at frequencies $\omega_s$ and $\omega_i$ are created. Conservation of energy dictates that $\omega_p=\omega_s+\omega_i$, while for macroscopic amplification and practical generation of coherent radiation at the signal and idler frequencies the phase-matching condition has to be satisfied
\begin{equation}
    \dk=k_p-k_s-k_i = 0,
\end{equation}
where $k_x=n(\omega_x)\omega_x/c$ ($x=p$,~$s$,~$i$) is the momentum at the frequency, $x$, with $n(\omega_x)$ the refractive index. The equations can generally be written as
\begin{numcases}{}
    \frac{\partial A_{p}^{(m)}}{\partial z} = \hat{L}_p A_{p}^{(m)} + i\kappa_p A_{s}^{(m)} A_{i}^{(m)}e^{-i\Delta k z} \label{eq:CEp}\\
    \frac{\partial A_{s}^{(m)}}{\partial z} = \hat{L}_s A_{s}^{(m)} + i\kappa_s A_{p}^{(m)} A_{i}^{*(m)}e^{+i\Delta k z} \label{eq:CEs}\\
    \frac{\partial A_{i}^{(m)}}{\partial z} = \hat{L}_i A_{i}^{(m)} + i\kappa_i A_{p}^{(m)} A_{s}^{*(m)}e^{+i\Delta k z}, \label{eq:CEi}
\end{numcases}
where the superscript $(m)$ is the number of the round-trip, $z$ is the spatial coordinate along the length of the crystal, $\kappa_x=2\pi\deff/n_x\lambda_x$ is the nonlinear coupling coefficient, and $\deff$ is the effective second-order susceptibility of the crystal. Here, $L_x$ are the linear operators
\begin{equation}\label{eq:linearop}
    \hat{L}_x = -\left[ \frac{\alpha_{x}}{2}+ \left(\frac{1}{\nu_s} - \frac{1}{\nu_x}\right) \frac{\partial}{\partial \tau}+i\frac{k^{''}_x}{2}\frac{\partial^2}{\partial \tau^2} + i\frac{k^{'''}_x}{3}\frac{\partial^3}{\partial \tau^3} \right],
\end{equation}
indicating that the CWEs are written in a co-moving frame at the signal frequency in the presence of linear attenuation, $\alpha_x$, group-velocities, $\nu_x$, group-velocity dispersion (GVD), $k^{''}_x$, and third-order dispersion (TOD), $k^{'''}_x$. It is to be noted that the incorporation of higher-order dispersion terms beyond the currently supported TOD is straightforward. \edited{Additionally, in our model we assume plane-wave approximation. Hence, diffraction terms are neglected in Eqs.~\ref{eq:CEp}-\ref{eq:CEi}.}
\vspace{0.5cm}
\noindent
\\ 
\textit{Boundary conditions:}\\
\noindent
After a single pass in the nonlinear crystal, every field must be refreshed before starting the next round-trip ($m+1$), if and only if that electric field is resonant in the optical cavity. This should take into account the intracavity losses and the cavity detuning given by \begin{equation}
      A_x^{(m+1)} = \sqrt{1-\theta_x}A_x^{(m)}e^{i\delta_x}, \label{eq:BC}
\end{equation}
where $\theta_x$ is the power transmission coefficient, $\delta_x=(\omega_{\mathrm{cav}}-\omega_x)\trt$ is the cavity detuning, $\omega_{\mathrm{cav}}$ is the frequency of a resonant mode, and $\trt$ is the round-trip time.

In what follows, we present the main features of our numerical
implementation of the CWEs in the form of the aforementioned equations, including the functionality to solve for nanosecond as well as cw operating regimes. The provided \verb|CUDA| package derived from our research software, is scripted with the aim not only to speed up the calculations, but also to keep it simple and adaptable to simulate other nonlinear processes. For example, the user can amend linear and nonlinear terms in Eqs.~\ref{eq:CEp}-\ref{eq:CEi}, add passive and/or active intracavity elements such as modulators, incorporate additional nonlinear crystals, and more.

\subsection{Split-step Fourier method}
\label{sec:ssfm}
\noindent

The SSFM is used here to model the propagation along the nonlinear medium, as schematically shown in Fig.~\ref{fig:scheme}(b). The crystal with length, $\lcr$, is discretized along the $z-$direction into steps of length, $dz$. In every step, the SSFM simultaneously solves the linear and nonlinear effects. The linear part is solved in the frequency domain, and the nonlinear part is solved in the time domain using a four-order Runge-Kutta method. This sequence is repeated throughout the length of the crystal. Depending on the implementation, this algorithm exhibits an error, $\mathcal{O}(dz^2)$ or $\mathcal{O}(dz^3)$. This is sequentially solved along the entire crystal and requires many operations with complex vectors as well as discrete Fourier transforms (DFTs) during the simulation, and depending on the complexity of the problem (i.e. vectors size, number of round trips inside the optical cavity, number of nonlinear crystals) this might be computationally demanding. 
 Equations~\ref{eq:CEp},~\ref{eq:CEs}, and~\ref{eq:CEi} can be written in the matrix form as (omitting the superscripts $m$)
\begin{equation}\label{eq:CWEsmatrix}
    \frac{\partial}{\partial z} 
    \begin{pmatrix}
    A_p\\
    A_s\\
    A_i
    \end{pmatrix} = 
    \underbrace{\begin{pmatrix}
    \hat{L}_p & 0 & 0  \\
    0 & \hat{L}_s & 0  \\
    0 & 0 & \hat{L}_i 
    \end{pmatrix}}_{\text{Linear operator}\\ \hat{L}}   
    \begin{pmatrix}
    A_p\\
    A_s\\
    A_i
    \end{pmatrix}
    +
    \underbrace{\begin{pmatrix}
    0 & 0 & \hat{N}_p \\
    \hat{N}_s & 0 & 0 \\
    \hat{N}_i & 0 & 0  
    \end{pmatrix}}_{\text{Nonlinear operator~}\\ \hat{N}}   
    \begin{pmatrix}
    A_p\\
    A_s\\
    A_i
    \end{pmatrix}
\end{equation}
where $\hat{L}_x$ is given by Eq.~\ref{eq:linearop}, and $\hat{N}_x$ are the corresponding nonlinear operators for each wavelenght, $\hat{N}_p = i\kappa_p A_s e^{-i\dk z}$, $\hat{N}_s = i\kappa_s A_i^* e^{i\dk z}$, and  $\hat{N}_i = i\kappa_i A_s^* e^{i\dk z}$, respectively. Equation~\ref{eq:CWEsmatrix} is then reduced to
\begin{equation}
    \frac{\partial}{\partial z} \Vec{A} = \left( \hat{L} + \hat{N} \right) \Vec{A}
\end{equation}
with a symbolic solution given by
\begin{equation}\label{eq:evol}
    \Vec{A}(z+dz) = e^{\left( \hat{L} + \hat{N} \right)dz} \Vec{A}(z)
\end{equation}

Since the operators, $\hat{L}$ and $\hat{N}$, in general do not commute, the approximation $e^{\left( \hat{L} + \hat{N} \right)dz} \approx e^{ \hat{L}dz} e^{\hat{N} dz}$, that yields an error, $\mathcal{O}(dz^2)$, is often used. 

However, in this work we implement a more accurate expression~\citep{agrawal2000nonlinear}
\begin{equation}\label{eq:expon}
    e^{\left( \hat{L} + \hat{N} \right)dz} \approx e^{ \hat{N}\frac{dz}{2}}e^{ \hat{L}dz} e^{ \hat{N}\frac{dz}{2}},
\end{equation}
with an error of $\mathcal{O}(dz^3)$. In this scheme, every step is solved by computing the nonlinear term in the first half-step, $dz/2$. After one Fourier transform, the linear term is computed in the entire step, $dz$. Finally, the nonlinear term is again computed in the second half-step, $dz/2$. This sequence, $\hat{N}/2-\hat{L}-\hat{N}/2$, is equivalent to its counterpart, $\hat{L}/2-\hat{N}-\hat{L}/2$, since both lead to the same solution. By inserting Eq.~\ref{eq:expon} in Eq.~\ref{eq:evol} and solving the linear part in the frequency domain, the field evolution reads
\begin{equation}
    \Vec{A}(z+dz) \approx e^{ \hat{N}\frac{dz}{2}} \invfourier{ e^{ \hat{L}dz} \fourier{e^{ \hat{N}\frac{dz}{2}}\Vec{A}(z)}} ,
\end{equation}
where $\fourier{\cdot}$ stands for the Fourier transform.

\subsection{Intracavity elements}
\label{sec:intracav}
\noindent

In our package, we include intracavity elements that are capable of being described mathematically. The user can incorporate any other intra- or extracavity element by deriving its proper expression. Here we include two intracavity elements that are usually employed in experiments:
\begin{enumerate}
    \item \textit{Dispersion compensation}: It is common in laser technology to compensate the GVD by including, e.g., chirped mirrors in the cavity. In the case of the OPO, the nonlinear crystal has an intrinsic GVD, as presented in the Sec.~\ref{sec:theory}. The use of chirped mirrors imposes an additional phase to the reflecting electric fields, such as
    $$\tilde{A}(\Omega)\rightarrow\tilde{A}(\Omega)e^{i\frac{\gamma}{2}\Omega^2},$$
    where the factor, $\gamma$, is related with the group-delay dispersion (GDD) and accountss for how much GVD is compensated, and $\Omega$ is the frequency. This operation must be performed in the frequency domain, $\tilde{A}$, so that an extra Fourier transform is required.
    \item \textit{Electro-optical modulator (EOM)}: This element is often used to broaden the spectrum of a given electric field. Physically, an intracavity EOM modulates the phase of an electric fields. Mathematically, the transformation in the time domain is given by 
    $$A(\tau)\rightarrow A(\tau)e^{i\beta\sin(2\pi\fpm \tau )},$$
    where $\fpm$ is the frequency of the EOM, typically on the order of the cavity free-spectral range, and $\beta$ the so-called modulation depth.    
\end{enumerate}

\subsection{Vector size and grid discretization considerations}
\label{sec:vecsizeconsid}
\noindent

\edited{The choice of the number of vector elements, as well as the step size of the spatial and temporal-spectral grids will depend on each specific problem. Here we simply provide a guide on what to consider when tackling the problem to be modeled.}\\

\edited{\underline{Spatial resolution, $dz$:}}

\edited{As described in Section~\ref{sec:ssfm}, the numerical error of this algorithm is $\mathcal{O}(dz^3)$, so decreasing the step size, $dz$,  the approximation of the numerical solution will be closer to the analytical solution. Typical values for the number of grid points at coordinate $z$ can range from 50 to 250~\citep{smith1999numerical}. One of the problems that can arise when the value of $dz$ is not small enough is that the code returns} \verb|NaN| \edited{values. This can be seen in the nonlinear part of the coupled equations. For simplicity, consider the case of the dispersionless degenerate OPO with perfect phase-matching ($\Delta k = 0$) and focus on the equation for the signal that can be approximated by}
\begin{equation*}
\frac{\partial A_{s}}{\partial z} = i\kappa_s A_{p} A_{s}^{*} \Rightarrow A_s(z+dz) \approx A_s(z) + i \Delta A_s(z),
\end{equation*}
\edited{where $\Delta A_s(z) = \left[\kappa_s A_p(z) dz \right] A_s^*(z)$ is the incremental change in the signal electric field, with $\kappa_s = 2\pi \deff/n_s\lambda_s$. With this approach, we expect the change in the electric field to be incremental, that is}
\begin{equation*}
\abs{\Delta A_s(z)} \ll \abs{A_s(z)} \approx \abs{A_s(z+dz)},    
\end{equation*}
\edited{in order to avoid undesired drastic changes. Since $\kappa_s$ and $A_p$ are experimental parameters, the only way to prevent any divergence is by varying $dz$. For example, for high pump powers or for crystals with a high nonlinear coefficient, $\deff$, it will be necessary to pay attention to the set value of $dz$.}\\

\edited{\underline{Temporal-spectral resolution, $dt-d\nu$:}}

\edited{It is worth noting that both step sizes are linked through the number of points, $n$, and the round-trip time, $\trt$, through}
\begin{equation*}
dt = \trt/n\text{~and~} d\nu= \trt^{-1}.    
\end{equation*}
\edited{Once the cavity round-trip time is calculated from the properties of the nonlinear crystal, it remains to determine the physics that the user aims to capture. In the case of the illustrative examples presented in the following sections (e.g., Fig.~\ref{fig:cw2eq} (c)), the relevant physics in the time domain exhibits oscillations of $\sim 1$~ps for the case of the chosen cavity, with $\trt\approx 100$~ps. To cover an oscillation of $\sim 1$~ps duration with 10 points, it requires $dt\approx 0.1$~ps, so that $n=\trt/dt \approx 1000$. This represents, in powers of 2, $n=2^{10}$. However, when investigating other complicated scenarios where intracavity elements are included, and there may be faster oscillations due to increased bandwidth, it is necessary to increase the number of points. For instance, in Ref.~\citep{Sanchez2022Ultrashort}, the oscillations obtained were sub-picosecond, requiring an increase in the number of points to $n=2^{14}$. Furthermore, there are OPO configurations for which stationary solutions are pulses with duration of $\sim 0.1$~ps in a round-trip time, $\trt\approx 100$~ps. In this scenario, additional points will help capture an accurate pulse structure. In such a case, a time step equal to or better than $dt=0.005$~ps requires the number of points to be $n=2^{14}$ or even $2^{15}$. Under these conditions, the} \verb|cuOPO| \edited{package can be exploited to explore the fast dynamics of complex OPOs.}

\section{Package description}
\label{sec:softdesc}
\noindent
The package, \verb|cuOPO|, was scripted in \verb|CUDA| programming language to run efficiently on a GPU. Its functionality depends on \verb|cuda-toolkits|~\citep{NVT}. The package was widely used in our group to model several scenarios related to OPOs, and tested on a Linux system. All the simulations were performed in a desktop computer using a microprocessor Intel(R) Core(TM) i7-9700 CPU @ 3.00GHz and a GPU NVIDIA GeForce GTX 1650. The package contains a main file, a folder with eight header files and a bash file that allows the user to compile and execute the package by enabling or disabling intracavity elements, the chosen nonlinear crystal, pumping regime, and others relevant parameters.

\subsection{Main file}
\label{sec:mainfile}

The main file, called \verb|cuOPO.cu|, is divided into seven main parts:
\begin{itemize}
    \item \textit{Setting GPU and timing}: Sets the intended GPU and starts the simulation timing.
	\item \textit{Define simulation parameters, physical quantities and set electric fields}: Defines all the simulation constants, namely, crystal, cavity, and fields parameters. We define the single-precision data types, \verb|real_t| and \verb|complex_t| (\verb|float| and \verb|cufftComplex|, respectively), that are needed to define scalars and vectors.
	\item \textit{Define GPU vectors}: Declares all the vectors needed to run SSFM in GPU.
	\item \textit{Main loop. Fields in the cavity}: Runs the core of the package. The function, \verb|EvolutionInCrystal(parameters)|, computes the single-pass electric fields, and it is embedded in a \verb|for-loop| that accounts for every round-trip. This function belongs to the header files, \verb|cwes2.h| and \verb|cwes3.h|, which contain all the required functions to perform the SSFM.
	\item \textit{Saving results}: Saves the simulation outcome.
	\item \textit{De-allocating memory from CPU and GPU}: Frees up the memory.
	\item \textit{Finish timing}: Finishes and returns the simulation runtime.
\end{itemize}

The user can also find the functions description in the source code and the full code overview in the \verb|README.md| file in the corresponding repository~\citep{repo}.

\subsection{Header files}
\noindent

The package contains eight header files in the folder \verb|headers|, which can be either modified or adapted to the user specific applications. 
\begin{itemize}
    \item \verb|common.h| contains functions necessary to check other functions executed on the GPU.
    \item \verb|operators.h| contains overloaded operators ($+,-,*,/$) to perform operations with complex numbers.
    \item \verb|functions.h| contains functions related to the initialization of the interacting electric fields and the incorporation of intracavity elements, among others.
    \item \verb|ppln.h|/\verb|spplt.h| contains the Sellmeier equations for two widely used nonlinear crystals (MgO:PPLN and MgO:sPPLT). For new crystals, user can create a new file with the corresponding refractive index data.
    \item \verb|files.h| contains four functions useful to save real or complex vectors into a \verb|.dat| file.
    \item \verb|cwes2.h|/\verb|cwes3.h| contain the functions needed to solve the Eqs.~\ref{eq:CEp}-\ref{eq:CEi} using the SSFM, depending on whether two or three CWEs are used. The user should properly modify the function, \verb|dAdz(parameters)|, according to the specific process to simulate.
\end{itemize}

\subsection{Compilation and execution}
\label{sec:softfunctionalities}
\noindent
Before running the code, it is necessary to compile the package and obtain an executable file. To do this, we execute the bash file, \verb|cuOPO.sh|, included in the package, which in turn contains the compilation and the execution command lines. The compilation command line is shown in Listing~\ref{lst:compilation}, where the compiler, \verb|nvcc|, is invoked to compile the file, \verb|cuOPO.cu|.

\begin{lstlisting}[caption={Compilation},label={lst:compilation},language=bash, basicstyle=\ttfamily,
  showstringspaces=false,
  commentstyle=\color{red},
  keywordstyle=\color{blue},
  basicstyle=\ttfamily\footnotesize,]
    # compile with the preprocessor variables and flags
    nvcc cuOPO.cu -D<REGIME> -D<CRYSTAL> -DTHREE_EQS
    --gpu-architecture=sm_75 -lcufftw -lcufft -o cuOPO
\end{lstlisting}
Notice there are three preprocessor variables, \verb|-D<...>|, required to compile the code, namely
\begin{itemize}
    \item \verb|-D<REGIME>| set the pumping regime, cw (\verb|-DCW_OPO|) or pulsed nanosecond (\verb|-DNS_OPO|) regime. It is mandatory to define this variable.
    \item \verb|-DTHREE_EQS| set the use of three CWEs. Do not declare this flag if only two CWEs are required (two CWEs is by default). Two CWEs are used at degeneracy, i.e., in SHG or degenerate parametric down-conversion.
    \item  \verb|-D<CRYSTAL>| allow to set between the two nonlinear crystals included, MgO:PPLN and MgO:sPPLT, by setting as \verb|-DPPLN| or \verb|-DSPPLT|, respectively. It is mandatory to define this variable.
\end{itemize}
As can be seen, there are some extra flags required for the compilation. The flag, \verb|--gpu-architecture=sm_75|, tells the compiler to use the specific GPU architecture. It is important to check the proper value for this flag according to the used GPU card. The flags, \verb|-lcufftw| and \verb|-lcufft|, are used for the Fourier transforms performed by \verb|CUDA|.

After successfully compiling, the next step is to run the code. In the file, \verb|cuOPO.sh|, there are some variables that will be passed as an argument to the main file, \verb|cuOPO.cu|. This, of course, can be modified by users who prefer just set the variables values in the main file. However, this may be useful when users need to systematically vary a physical quantity, e.g. pumping level, pump wavelength, turn on/off some intracavity element, etc. The execution line is shown in Listing~\ref{lst:execution}.

\begin{lstlisting}[caption={Execution},label={lst:execution},language=bash, basicstyle=\ttfamily,
  showstringspaces=false,
  commentstyle=\color{red},
  keywordstyle=\color{blue},
  basicstyle=\ttfamily\footnotesize,]
    # exectute with the passed arguments
    ./cuOPO $<SET_OF_ARGUMENTS_TO_PASS>
\end{lstlisting}

Once the execution is finished, the output files are moved to a specific folder created by the file, \verb|cuOPO.sh|. The name of the created folder is related to the simulation parameters, but this can be changed according to the user requirements.

\subsection{Performance GPU vs. CPU}
\label{sec:performance}
\noindent

The operations performed by SSFM are essentially sums, products, and discrete Fourier transforms (DFTs) of complex vectors. The use of a GPU is substantially justified for sufficiently large vector size, $n$. The ratio of the time for a given calculation performed on CPU to that obtained in GPU is called \textit{speedup}, and is a way to measure the performance of the GPU scripted algorithm. \edited{To measure the speedup we compare the execution time of our SSFM implementation scripted in CUDA and in C language.}
The performed DFTs on CPU were calculated with the widely used FFTW library. The execution of DFT on GPU was carried out using \verb|cuFFT|, the CUDA FFT. Both DFT implementations have an order of convergence, $\mathcal{O}(n \log n)$~\citep{frigo2005design, cuFFT}. On the other hand, the rest of operations sum and products, have an order of convergence, $\mathcal{O}(n)$. The global algorithm has a convergence order dominated by the DFT. However, as the number of round-trips increases ($\sim 10^4$), the order of convergence might be closer to $\mathcal{O}(n^2)$.

\begin{figure}[htbp]
    \centering
    \includegraphics[width=1.0\textwidth]{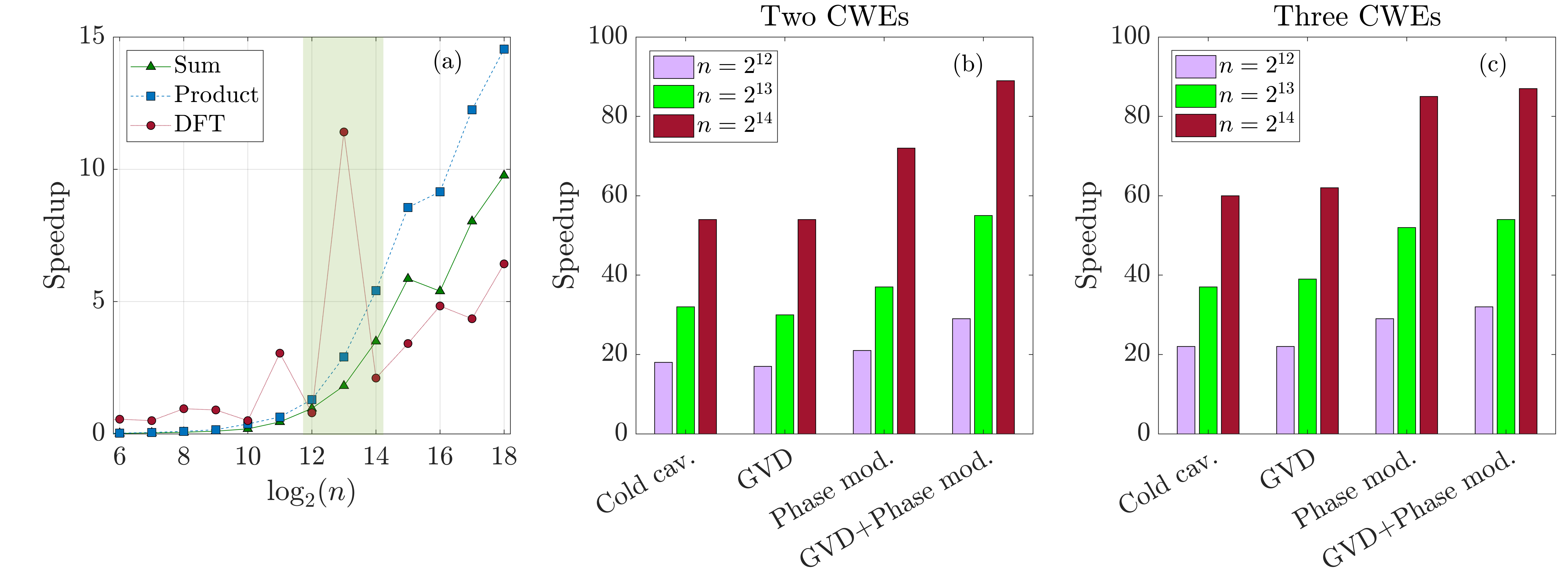}
    \caption{Performance study of the implementation. (a) Complex operations speedup as a function of the vectors size, $n$. The green shaded region corresponds to the interesting vector size of $2^{12},2^{13},~2^{14}$ elements. Speedup in different OPO configurations for (b) two and (c) three CWEs.}
    \label{fig:speedup}
\end{figure}

Figure~\ref{fig:speedup}(a) shows the speedup of the complex sums, products, and DFTs for single-precision operations. As expected, as the vector size increases the speedup also increases. The green shaded region emphasises the vector sizes used for the testing simulations that are typically sufficient to model the temporal and spectral behavior of OPOs. Since the obtained speedup is $>1$ for all the operations and for vector size $\geq 2^{12}$, the use of  GPU is justified. The performance of two CWEs (Fig.~\ref{fig:speedup}(b)) and three CWEs (Fig.~\ref{fig:speedup}(c)) is shown, for different OPO configurations, for three vector sizes. We refer to OPO configurations the set of components present in addition to the nonlinear crystal and mirrors, such as intracavitary elements. The first configuration is called \comillas{cold cavity} in which there is no intracavity elements (only a single pass is computed). The second configuration includes dispersion compensation by using chirped mirrors. The third configuration includes an intracavity EOM. Finally, the fourth configuration includes both dispersion compensation and EOM. As can be clearly verified, the speedup exceeds the value of 50 for $n=2^{14}$, greatly justifying the use of the implementation in GPU. For instance, to simulate $10^4$ round-trips in the cw regime using two CWEs in an OPO configuration with GVD compensation and intracavity EOM, the runtime simulation is around 30 seconds/500 round-trips on GPU, while in CPU results in 45 minutes/500 round-trips. However, the cw simulations requires thousands of round-trips to converge the steady-state solutions.

\section{Illustrative Examples}
\label{sec:examples}
\noindent
As illustrative examples, we present the use of \verb|cuOPO| package in different time scales (cw or pulsed nanosecond) for various TWM processes. As mentioned in Sec.~\ref{sec:softfunctionalities}, it is necessary to use a preprocessor variable in the compilation command line to select the regime. As also noted, there are only two values that this variable can take: \verb|-DCW_OPO| or \verb|-DNS_OPO|, and it is mandatory to define one of them during compilation. 

In each example, two listings are shown: the first contains the command line needed for the compilation, depending on each case; the second contains the specific lines for each example, but not all the code due to its large size. In the latter, it is mainly described how to initialize the electric fields, declare resonant and non-resonant fields in the cavity, and how to save the complex electric fields in a \verb|.dat| file.

\subsection{Main loop: The ring cavity}
\label{sec:mainloop}
\noindent
Before starting with the illustrative examples, it is recommended to surf through the code and have an overview of its structure. Listing~\ref{lst:mainloop} shows the main loop in which the fields propagate inside the cavity over several round-trips (line 447 in file \verb|cuOPO.cu|). The first is the evolution along the nonlinear crystal, using the overloaded function, \verb|EvolutionInCrystal( parameters )|, where the electric fields are calculated using the CWEs. Additional effects such as a phase due to detuning through $\delta$, cavity losses through $\sqrt{R}$, incorporation of intracavity elements such as an EOM, or dispersion compensation, are then implemented. Once this process is concluded, the next round-trip is calculated, where the previously calculated electric fields are now the input fields. This process is repeated for the number of round-trips, \verb|NRT|, set by the user as a global constant at the beginning of the main file. The functions
\begin{itemize}
    \item \verb|ReadPump<<<...>>>(params.)|: Reads the input pump and copies to \verb|Ap_gpu|, the vector used as intracavity pump field.
    \item \verb|AddGDD<<<...>>>(params.)|: Adds an additional phase in the frequency domain, useful to compensate the net cavity dispersion.
    \item \verb|PhaseModulatorIntraCavity<<<...>>>(params.)|: This function incorporates an EOM as a intracavity element. The user can incorporate any intracavity element just by knowing its functional form.
    \item \verb|SaveRoundTrip<<<...>>>(params.)|: save the current round-trip into the output vector, which will later be saved in a \verb|.dat| file.
\end{itemize}
are the so-called \textit{CUDA kernels} that are executed on the GPU.
\begin{lstlisting}[caption={Structure of the part of the code corresponding to the cavity simulation.}, label={lst:mainloop}, language=C++, basicstyle=\ttfamily,
  showstringspaces=false,
  commentstyle=\color{red},
  keywordstyle=\color{blue},
  basicstyle=\ttfamily\footnotesize,
  tabsize=2,]
// Main loop (fields in the cavity)
for (int nn = 0; nn < NRT; nn++){
    ...
    // This function compute the CWEs performing the SSFM
    EvolutionInCrystal(parameters);
	
    // Dispersion compensation
    if(GDD!=0){
        AddGDD<<<grid,block>>>(parameters);
        cufftExecC2C(plan1D, (complex_t *)Asw_gpu,
            (complex_t *)As_gpu, CUFFT_FORWARD);
    }		
	
    // If As is resonant, adds phase and losses.
    if (is_As_resonant){
        AddPhase<<<grid,block>>>(parameters);
    }
    
    // Use an intracavy phase modulator
    if( using_phase_modulator ){
        PhaseModulatorIntraCavity<<<grid,block>>>(parameters);
    }
    ...
}
\end{lstlisting}

\subsection{Nanosecond input pulse in an OPO}
\label{sec:nsopo}
\begin{figure}[ht]
    \centering
    \includegraphics[width=0.6\textwidth]{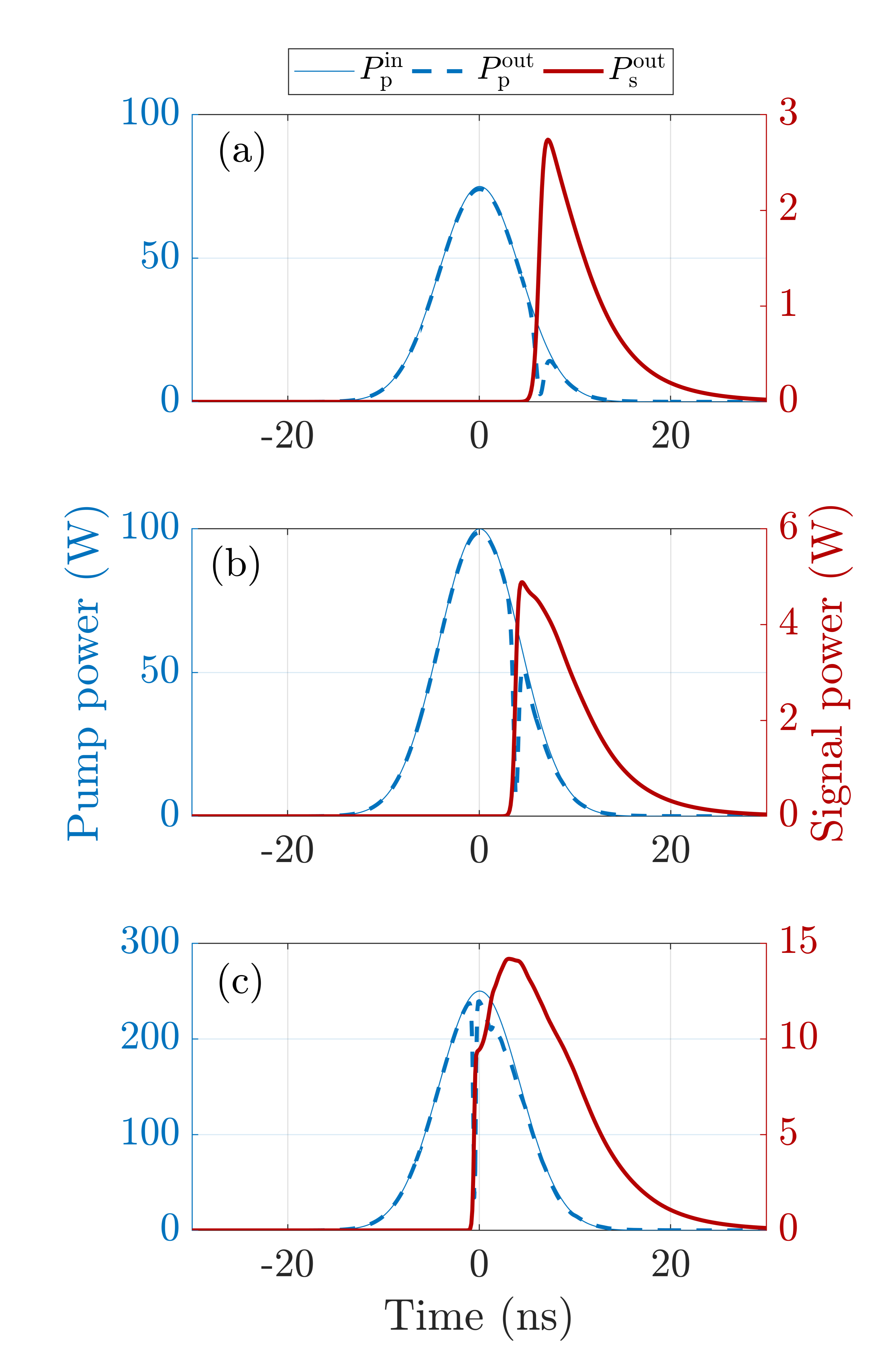}
    \caption{Nanosecond pump regime for a peak power of (a) 75~W, (b) 100~W, and (c) 250~W. The pump and signal wavelengths are $\lambda_p=532$~nm and $\lambda_p=1064$~nm, respectively. Blue curves correspond to the input (solid) and output (dashed) pump, whilst the red curve correspond to the extracavity output signal.} 
    \label{fig:nsexample}
\end{figure}

\noindent
The first example is shown in Fig.~\ref{fig:nsexample}. An input pulse with full-width at half-maximum (FWHM) duration of 10~ns and different peak powers of (a) 75~W, (b) 100~W, and (c) 250~W, is used as a pump field in the cavity. The pump wavelength is 532~nm and the signal wavelength is 1064~nm. This means that this OPO operates at degeneracy. Here and in the next examples, we use a 5-mm-long nonlinear crystal.
Figure~\ref{fig:nsexample} also shows the output pump and the generated signal. As expected, the greater the power of the pump, the greater its depletion and faster the signal rise time. 

Because this OPO operates at degeneracy, we only need two CWEs, since signal and idler are indistinguishable. Thus, for nanosecond regime and for only two CWEs, the compilation line is shown in Listing~\ref{lst:compns2eq}. This compilation line should be accordingly modified in line 23 or 26 in the \verb|cuOPO.sh| file.
\begin{lstlisting}[caption={Compilation for nanosecond regime and two CWEs.},label={lst:compns2eq},language=bash, basicstyle=\ttfamily,
  showstringspaces=false,
  commentstyle=\color{red},
  keywordstyle=\color{blue},
  basicstyle=\ttfamily\footnotesize,]
nvcc cuOPO.cu -DNS_OPO -DPPLN --gpu-architecture=sm_75
    -lcufftw -lcufft -o cuOPO
\end{lstlisting}

Listing~\ref{lst:ns_resonant} summarizes the important variables to set the input pump and signal fields, set the resonant fields to make singly-, doubly-, or triply-resonant oscillator (SRO, DRO and TRO, respectively) cavity, and how to save the results to a \verb|.dat| file using the function \verb|SaveVectorComplexGPU|.
\begin{lstlisting}[caption={Set the conditions for simulate the results in Fig.~\ref{fig:nsexample}. These portions of code are in the main file cuOPO.cu.}, label={lst:ns_resonant}, language=C++, basicstyle=\ttfamily,
  showstringspaces=false,
  commentstyle=\color{red},
  keywordstyle=\color{blue},
  basicstyle=\ttfamily\footnotesize,]
// Difine which fields are resonant (SRO, DRO or TRO)
bool is_Ap_resonant = false;
bool is_As_resonant = true;
#ifdef THREE_EQS // this is not executed (two CWEs)
bool is_Ai_resonant = true;
#endif
.
.
#ifdef NS_OPO // For nanosecond regime
real_t FWHM = 10000; // intensity FWHM for input [ps]

// set input pump (temporal Gaussian profile)
complex_t *Ap_in=(complex_t*)malloc(sizeof(complex_t)*SIZEL);
input_field_T(Ap_in, Ap0, Tp, sigmap, SIZEL); 
#endif
.
.
// Define input signal vector (NOISE)
complex_t *As = (complex_t*)malloc(nBytes);
NoiseGeneratorCPU (As, SIZE);
.
.
// Save data to .dat files
Filename = "signal_output";	
SaveVectorComplexGPU (As_total, SIZEL, Filename);
Filename = "pump_output";
SaveVectorComplexGPU (Ap_total, SIZEL, Filename);
\end{lstlisting}

The pump input field, \verb|Ap_in|, is initialized using the overloaded function, \verb|input_field_T(Ap_in, Ap0, Tp, sigmap, SIZEL)|, where \verb|Ap0| is the electric field strength related to the pump power, \verb|Tp| is the total time vector, \verb|sigmap| is standard deviation of the Gaussian pulse, and \verb|SIZEL| is the vector size. The initial signal electric field starts from random amplitude and phase, using the function \verb|NoiseGeneratorCPU(As, SIZE)|. It is important to note that in our implementation, the pump power is passed as an external parameter through the shell script, \verb|cuOPO.sh|. This is because for these examples we perform successive simulations by varying the input power.

\subsection{Continuous-wave OPO using 3 CWEs}
\label{sec:cw3opo}
\begin{figure}[htbp]
    \centering
    \includegraphics[width=1.0\textwidth]{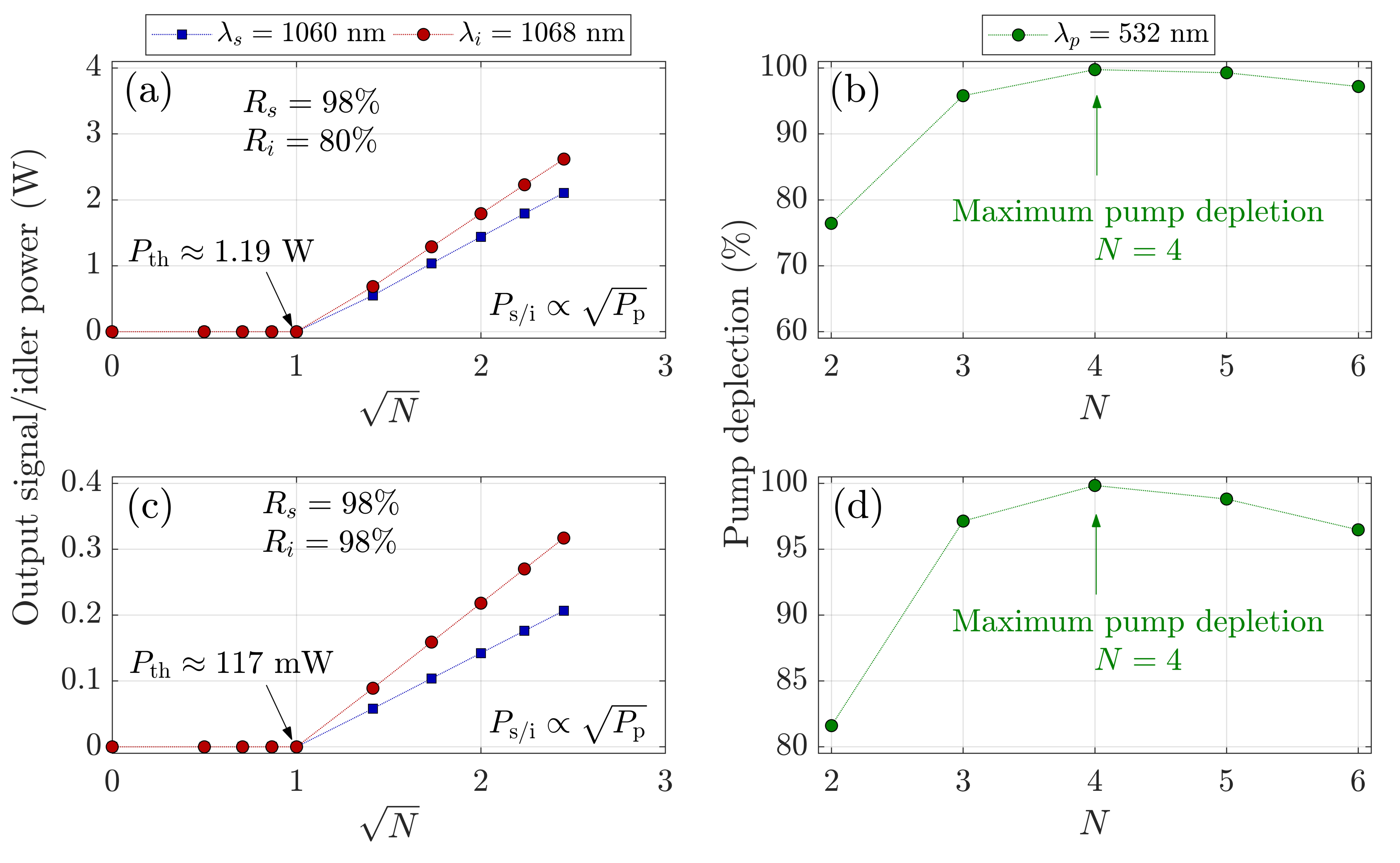}
    \caption{Simulations in the cw regime and three CWEs. The panels show the dependency of the signal/idler power as a function of the pump power, for a pump wavelength at $\lambda_p=532$~nm. The signal/idler reflectivities are the same (a) and different (b).}
    \label{fig:cw3example}
\end{figure}

\noindent
In the second example, we model a doubly-resonant cw-pumped OPO at 532~nm to generate signal and idler at 1060~nm and 1068~nm, respectively. In this case, both signal and idler are resonant, while the pump is single-pass. The results are presented in terms of the pumping level, $N=\Pin/\Pth$, defined as the ratio of the input pump power to the threshold pump power, where~\citep{ebrahimzadeh2001optical}
\begin{equation}\label{eq:pth}
    \Pth = \frac{\epsilon_0cn_pn_sn_i\lambda_s\lambda_iw_{0p}^2}{8\pi\deff^2\lcr^2}\alpha_s\alpha_i,
\end{equation}
with $\epsilon_0$ is the vacuum dielectric constant, $c$ the speed of light, and $w_{0p}=55\micron$ the pump beam waist in the crystal. The total losses at signal and idler frequencies are $\alpha_x=(1-R_x + \alpha_{cx}\lcr)/2$ ($x=s,i$), where $R_x$ and $\alpha_{cx}$ are the corresponding reflectivity and attenuation. 
The simulation results are shown in Fig.~\ref{fig:cw3example}, where the averaged output signal/idler power as a function of $\sqrt{N}$ (or equivalently, $\sqrt{P_{\mathrm{p}}}$) is depicted in panels (a) $(R_s,R_i)=(0.98,0.80)$ and (c) $(R_s,R_i)=(0.98,0.98)$. In panels (b) and (d) we plot the pump depletion, $(1-P_{\mathrm{p}}^{\mathrm{out}}/P_{\mathrm{p}}^{\mathrm{in}})\times 100\% $, as a function of $N$, clearly indicating that the maximum pump depletion is achieved at $N=4$, as expected~\citep{ebrahimzadeh2001optical}. Notice that the threshold condition ($\Pth\approx1.19$~W and $\Pth\approx117$~mW) is numerically computed for this doubly-resonant OPO.

The command line to compile the cw regime using three CWEs is shown in Linting~\ref{lst:compcw3eq}. Note that we added the preprocessor variables, \verb|-DTHREE_EQS|, to specify the correct amount of equations, and the variable, \verb|-DCW_OPO|, to set the regime and the variable, \verb|-DPPLN|, to set the crystal. This compilation line should be accordingly modify in line 23 or 26 in the \verb|cuOPO.sh| file.
\begin{lstlisting}[caption={Compilation for cw regime and three CWEs},label={lst:compcw3eq},language=bash, basicstyle=\ttfamily,
  showstringspaces=false,
  commentstyle=\color{red},
  keywordstyle=\color{blue},
  basicstyle=\ttfamily\footnotesize,]
nvcc cuOPO.cu -DCW_OPO -DPPLN -DTHREE_EQS --gpu-architecture=sm_75
    -lcufftw -lcufft -o cuOPO
\end{lstlisting}

\edited{Listing}~\ref{lst:cw3_resonant} shows the relevant setting to run these simulations. By defining the preprocessor variable, \verb|-DCW_OPO|, the code calls the overloaded function, \verb|input_field_T|, that initializes the input pump. As in the pulsed nanosecond regime, both signal and idler initial electric fields start from random amplitude and phase, using the function, \verb|NoiseGeneratorCPU(A,SIZE)|.

\begin{lstlisting}[caption={Set the conditions for simulate the results in Fig.~\ref{fig:cw3example}. These portions of code are in the main file cuOPO.cu.}, label={lst:cw3_resonant}, language=C++, basicstyle=\ttfamily,
  showstringspaces=false,
  commentstyle=\color{red},
  keywordstyle=\color{blue},
  basicstyle=\ttfamily\footnotesize,]
// Define which fields are resonant (SRO, DRO or TRO)
bool is_Ap_resonant = false;
bool is_As_resonant = true;
#ifdef THREE_EQS // this is now executed (three CWEs)
bool is_Ai_resonant = true;
#endif
.
.
// Set input pump in cw regime
#ifdef CW_OPO
complex_t *Ap_in = (complex_t*)malloc(nBytes); 
input_field_T(Ap_in, Ap0, SIZE);
#endif
.
.
// Define input signal vector (NOISE)
complex_t *As = (complex_t*)malloc(nBytes);
NoiseGeneratorCPU (As, SIZE);
// this line runs when set -DTHREE_EQS in the compilation
#ifdef THREE_EQS
// Define input idler vector (NOISE)
complex_t *Ai = (complex_t*)malloc(nBytes);
NoiseGeneratorCPU (Ai, SIZE);
#endif
.
.
// Save data to .dat files
Filename = "signal_output";	
SaveVectorComplexGPU(As_total, SIZEL, Filename);
Filename = "pump_output";
SaveVectorComplexGPU(Ap_total, SIZEL, Filename);
// this line runs when set -DTHREE_EQS in the compilation
#ifdef THREE_EQS
Filename = "idler_output";
SaveVectorComplexGPU(Ai_total, SIZEL, Filename);		
#endif
\end{lstlisting}

\subsection{Continuous-wave OPO using 2 CWEs}
\label{sec:cw2opo}
\begin{figure}[htbp]
    \centering
    \includegraphics[width=1.0\textwidth]{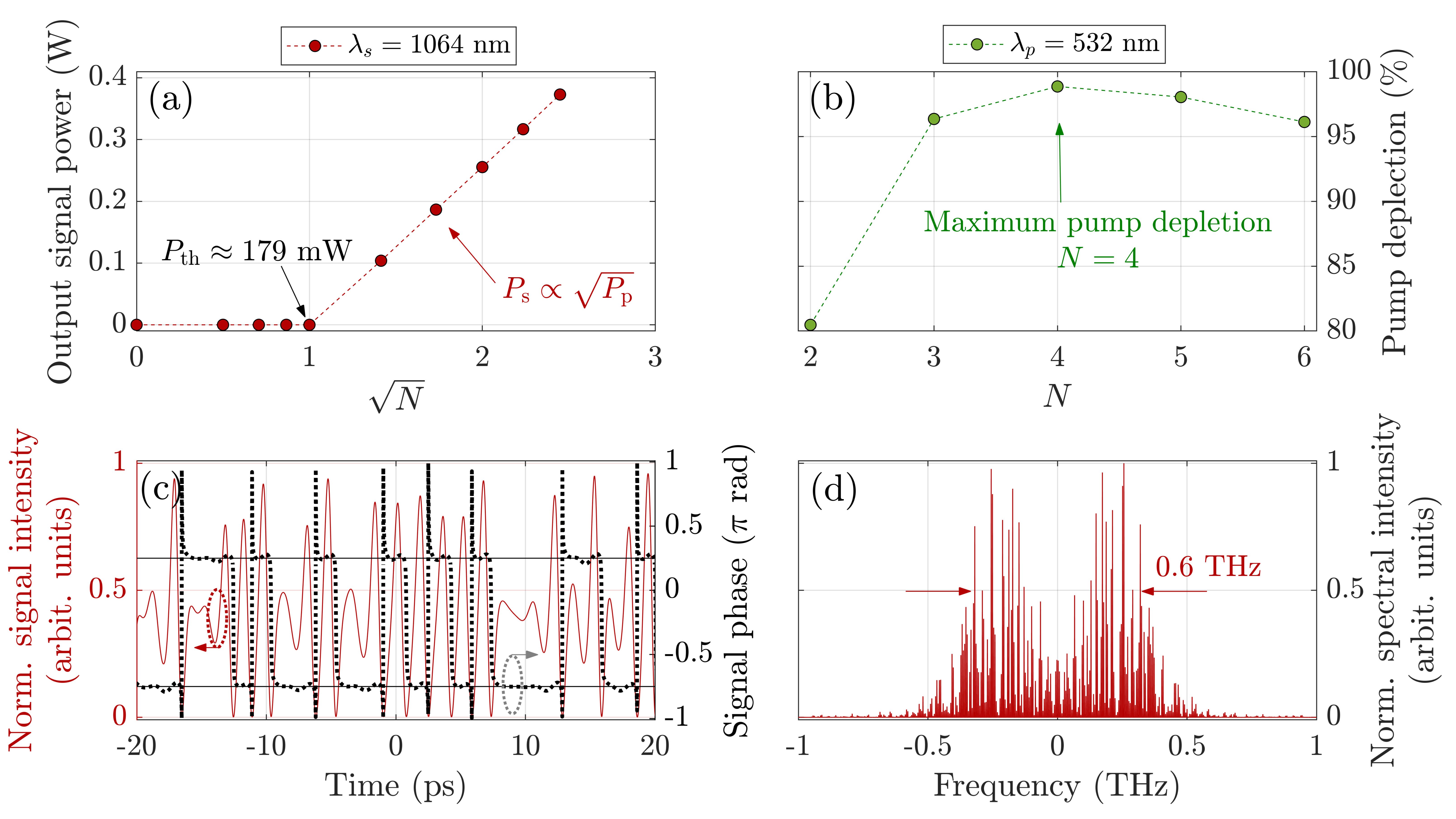}
    \caption{Simulations in the cw regime and two CWEs for an OPO working at degeneracy. Panel (a): dependency of the signal power as a function of the square root of the pumping level, $\sqrt{N}$, for a pump wavelength at $\lambda_p=532$~nm. Panel (b): pump depletion as a function of $N$. Panel (c):  left axis shows the normalized intensity as a function of the time; right axis shows the signal phase as a function of the time. Panel (d) the corresponding signal spectrum.}
    \label{fig:cw2eq}
\end{figure}

\noindent
The third example concerns the two CWEs in a DRO configuration, in which the OPO operates at degeneracy (signal and idler are not distinguishable), as shown in Sec.~\ref{sec:nsopo}. In this case, a pump wavelength at $\lambda_p=532$~nm generates a degenerate field at $\lambda_s=1064$~nm. As expected, the signal and pump power follow the same relation as in Sec.~\ref{sec:cw3opo}, that is $P_{\mathrm{s}}= \sqrt{P_{\mathrm{p}}}$. This is shown in Fig.~\ref{fig:cw2eq} (a), where after the threshold pumping level is breached the generated signal linearly scales with $\sqrt{N}$. Figure~\ref{fig:cw2eq} (b) shows the pump depletion as a function of $N$, which is again maximized at $N=4$. Figure.~\ref{fig:cw2eq} (c) shows a short-time slice of 40~ps (left axis) of the total round-trip time, $\trt\approx102$~ps, of the normalized signal intensity ($\propto \abs{A_s}^2$) and the corresponding phase, $\phi_s$ (right axis). As can be seen, the phase takes two possible values, $\pi/4$ and $-3\pi/4$, as expected from theoretical analisys~\citep{wong2010self}. The corresponding spectrum is shown in Fig.~\ref{fig:cw2eq} (d). 

The compilation command line using two CWEs is shown in \edited{Listing}~\ref{lst:compcw2eq}. Note that we omitted the preprocessor variable, \verb|-DTHREE_EQS|, since two CWEs is set as a default. Once again, the variable, \verb|-DCW_OPO|, is declared to set the cw regime and the variable, \verb|-DPPLN|, to set the crystal. This compilation line should be accordingly modified in line 23 or 26 in the \verb|cuOPO.sh| file.
\begin{lstlisting}[caption={Compilation for cw regime and two CWEs},label={lst:compcw2eq},language=bash, basicstyle=\ttfamily,
  showstringspaces=false,
  commentstyle=\color{red},
  keywordstyle=\color{blue},
  basicstyle=\ttfamily\footnotesize,]
nvcc cuOPO.cu -DCW_OPO -DPPLN --gpu-architecture=sm_75
    -lcufftw -lcufft -o cuOPO
\end{lstlisting}

\edited{Listing}~\ref{lst:cw2_resonant} shows the relevant setting to run these simulations. By defining the preprocessor variable, \verb|-DCW_OPO|, the code calls the overloaded function, \verb|input_field_T|, that initializes the input pump. As in the previous examples, signal initial electric field starts from random amplitude and phase, using the function, \verb|NoiseGeneratorCPU(As,SIZE)|.

\begin{lstlisting}[caption={Set the conditions for simulate the results in Fig.~\ref{fig:cw2eq}. These portions of code are in the main file cuOPO.cu.}, label={lst:cw2_resonant}, language=C++, basicstyle=\ttfamily,
  showstringspaces=false,
  commentstyle=\color{red},
  keywordstyle=\color{blue},
  basicstyle=\ttfamily\footnotesize,]
// Set input pump in cw regime
#ifdef CW_OPO
complex_t *Ap_in = (complex_t*)malloc(nBytes); 
input_field_T(Ap_in, Ap0, SIZE);
#endif
.
.
// Define which fields are resonant (SRO, DRO or TRO):
// single pass = false / resonant = true
bool is_Ap_resonant = false;
bool is_As_resonant = true;
.
.
// Define input signal vector (NOISE)
complex_t *As = (complex_t*)malloc(nBytes);
NoiseGeneratorCPU(As, SIZE);
.
.
// Save data to .dat files
Filename = "signal_output";	
SaveVectorComplexGPU(As_total, SIZEL, Filename);
Filename = "pump_output";
SaveVectorComplexGPU(Ap_total, SIZEL, Filename);
\end{lstlisting}

\section{Conclusions}
\label{sec:conclusions}
\noindent
In this work we have presented an efficient and fast software tool that allows us to study and numerically model optical parametric oscillators. This tool implements the symmetrized split-Step Fourier Method with the purpose of solving the coupled differential equations that describe the propagation of light in second-order nonlinear media represented by Eqs.~\ref{eq:CEp}-\ref{eq:CEi}. 

In order to speed up the simulations, which are often computationally demanding, the code was scripted in the CUDA programming language to be executed on a graphics card, or GPU. The achieved increase in speed will depend on the GPU used. Here, we obtained a speedup of 50X in comparison with an equivalent CPU-based implementation using a GPU GeForce GTX 1650 for vector size $n=2^{14}$.

The package allows the modeling of OPOs that operate in the pulsed nanosecond or continuous-wave regime in practical configurations, and the cavity can determine which fields will or will not be resonant depending on the particular scheme (SRO, DRO, TRO). A single nonlinear crystal was used in the implementation of this package, but the incorporation of one or more nonlinear media, as well as the addition of intracavity elements, is simple and straightforward.

The package returns as an output text files \verb|<output_file>.dat| containing the vectors of time, frequency, and electric fields. Since the electric field is a complex function in the time domain, we have decided to create a different file for the real and imaginary part separately. Finally, each user can use these outputs for post-processing, for example by reading these files in Python or Matlab.

We believe that the community will find this package useful for studying and modeling not only OPOs, but also other systems similar to those presented here, by adapting or modifying the routines deployed in \verb|cuOPO|.





\bibliographystyle{elsarticle-num}
\bibliography{main}







\end{document}